\documentclass[conference]{IEEEtran}
\usepackage{cite}

\usepackage{graphicx,epsfig,amssymb}

\ifCLASSINFOpdf
\else
\fi
\usepackage{threeparttable}
\usepackage{tabu}
\usepackage[cmex10]{amsmath}
\begin{document}
%
\title{Downlink Small-cell Base Station Cooperation Strategy in Fractal Small-cell Networks}



%
\author{\IEEEauthorblockN{Fen Bin\IEEEauthorrefmark{1},
Jiaqi Chen\IEEEauthorrefmark{1},
Xiaohu Ge\IEEEauthorrefmark{1},
and Wei Xiang\IEEEauthorrefmark{2},
}
\IEEEauthorblockA{\IEEEauthorrefmark{1}School of Electronic Information and Communications\\ Huazhong University of Science and Technology, China}
\IEEEauthorblockA{\IEEEauthorrefmark{2}James Cook University, Australia}
\IEEEauthorblockA{Corresponding author: Xiaohu Ge. Email:
\IEEEauthorrefmark{1}\{xhge\}@hust.edu.cn}}




\maketitle

\begin{abstract}
Coordinated multipoint (CoMP) communications are considered for the fifth-generation (5G) small-cell networks as a tool to improve the high data rates and the cell-edge throughput. The average achievable rates of the small-cell base stations (SBS) cooperation strategies with distance and received signal power constraints are respectively derived for the fractal small-cell networks based on the anisotropic path loss model. Simulation results are presented to show that the average achievable rate with the received signal power constraint is larger than the rate with a distance constraint considering the same number of cooperative SBSs. The average achievable rate with distance constraint decreases with the increase of the intensity of SBSs when the anisotropic path loss model is considered. What's more, the network energy efficiency of fractal small-cell networks adopting the SBS cooperation strategy with the received signal power constraint is analyzed. The network energy efficiency decreases with the increase of the intensity of SBSs which indicates a challenge on the deployment design for fractal small-cell networks.
\end{abstract}
\vspace{2 ex}



%
\IEEEpeerreviewmaketitle

\section{Introduction}
\label{sec1}
Greater requirements on data rates, the number of connected devices and network capacity are demanded of the fifth generation (5G) communication system\cite{V1}. To satisfy the requirements of the 5G communication system, dense deployment of cellular networks is inevitable in the future networks. A large number of low power base stations (BSs) are deployed in lieu of a traditional macro cell, which significantly increase network throughput and capacity\cite{T2}. On the other hand, interference deteriorates due to the dramatic increase in the number of interference sources. In this case, coordinated multipoint is proposed to avoid or exploit interference with the objective of improving the cell edge and average data rates. BS cooperation on the downlink can improve the average throughput and, more importantly, cell edge throughput. The user data to be transmitted to one user equipment (UE) is available in multiple BSs of the network, and is simultaneously transmitted from multiple BSs.

In the literature, cooperative BS techniques for traditional cellular networks have been well studied\cite{S2,X4,G5,W6}. A clustered multi-cell coordination in a cellular system with randomly deployed BSs is proposed in\cite{S2}, and the average achievable rate is derived for a typical user which indicates that the average achievable rate with interference coordination increases with the average cluster size. In \cite{X4}, the vehicular mobility performance is analyzed based on the distances between the vehicle and cooperative small-cell BSs (SBSs) for 5G cooperative multi-input multi-output (MIMO) small-cell networks in consideration of cochannel interference. An integral expression for the network coverage probability is derived in \cite{G5} considering SBS cooperation in downlink heterogeneous cellular networks, which has shown that SBS cooperation is more beneficial for the worst-performing user compared to the general population. The selection of cooperative SBSs is based on the distances between the desired user and SBSs in \cite{X4,G5}. Another common selection is based on the received signal strength (RSS) at the desired user. A user-centric clustering model, based on a tier-specific RSS threshold, is proposed in \cite{W6} for SBS cooperation in the downlink heterogeneous cellular networks, and a power minimization problem with a minimum spectral efficiency constraint is formulated. An approximate solution is derived to show its high accuracy via simulation results.

The works in \cite{S2,X4,G5,W6} assume that the path loss is isotropous in a cellular scenario or cellular tire. However, buildings and obstacles are distributed irregularly in urban environments, and electromagnetic waves of different directions experience different fading given diffraction and scattering effects in different propagation directions. Therefore, the path loss exponents differ not only in different propagation distance ranges, but also in different propagation directions even with the same distance range in practical cellular scenarios. The path loss is anisotropic in practical cellular scenarios. In this case, the anisotropic path loss is an inevitable challenge to investigate the SBS cooperation in small-cell networks. Based on the fractal characteristics of cellular coverage and the anisotropic path loss model in \cite{G8}, the average achievable rate and network energy efficiency adopting SBS cooperation strategies are derived to investigate the SBS cooperation performance in fractal small-cell networks.
The main contributions of this paper are three-fold:
\begin{enumerate}
\item Considering the fractal characteristic of cellular coverage, the anisotropic path loss model is proposed to analyze the SBS cooperation performance of random small-cell networks;
\item Based on the anisotropic path loss model, the average achievable rates of the SBS cooperation strategies with distance and received signal power constraints are derived for fractal small-cell networks. Compared with the cooperation strategy based on distances, numerical results indicate that the maximum average achievable rate with the cooperation strategy based on received signal power is improved by 700\%;
\item The network energy efficiency of fractal small-cell networks with received signal power constraint is derived based on the anisotropic path loss model. The numerical results show that the network energy efficiency decreases with the increase of the intensity of SBSs.
\end{enumerate}

The remainder of this paper is structured as follows. Section II describes the system model. The average achievable rates with SBS cooperation strategies are presented in Section III. The network energy efficiency of fractal small-cell networks is derived in Section IV, followed by simulation results discussed in Section V. Finally, conclusions  are drawn in Section VI.

\section{SYSTEM MODEL}
\label{sec2}
 \begin{figure}
 \setlength{\abovecaptionskip}{0.cm}
\setlength{\belowcaptionskip}{-0.cm}
  \centering
  \includegraphics[width=7cm,draft=false]{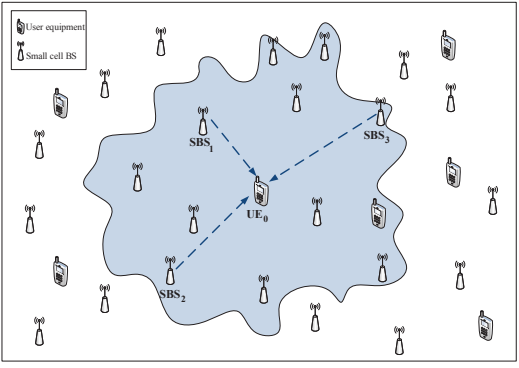}\\
  \caption{\small System model of the downlink small-cell base station cooperation in fractal small-cell networks. Dark region denotes the coverage area of a typical user which has a fractal boundary.}
  \label{fig1}
\end{figure}
In this paper, SBS cooperation in the downlink of 5G fractal small-cell networks is investigated. Assume that both SBSs and user equipment (UEs) are located randomly in the infinite plane $\mathbb{R}^2$. The locations of SBSs and UEs can be modeled as two independent homogeneous Poisson point processes (HPPP)\cite{A13,G1}, denoted by ${\Phi _B} = \left\{ {{x_i},i = 1,2,3,...} \right\}$ and ${\Phi _U} = \left\{ {{y_u},u = 1,2,3,...} \right\}$, where ${x_i}$ and ${y_u}$ are two-dimensional Cartesian coordinates, denoting the locations of the $i$-th SBS $SBS_i$ and the $u$-th UE $UE_u$, respectively. The corresponding intensities of the two Poisson point processes are ${\lambda _B}$ and ${\lambda _U}$.
To evaluate the received signal power at a typical UE, it is assumed that the UE is located at the coordinate origin $o$, denoting by UE$_0$. The received signal power at UE$_0$ from $SBS_i$ is expressed as

\begin{equation}
{P_i} = {P_s}{h_i}{L_i},
\label{eq1}
\tag{1}
\end{equation}
where $h_i$ refers to the Rayleigh fading between UE$_0$ and $SBS_i$, which is distributed as an exponential distribution with mean one\cite{A14}. For brevity but without loss of generality, all the SBSs are assumed to transmit with the same transmission power $P_s$. $L_i$ denotes the path loss between UE$_0$ and $SBS_i$. The system model of a 5G fractal small-cell networks is illustrated in Fig. 1.

The path loss between UE$_0$ and $SBS_i$ in the 5G fractal small-cell networks is expressed as
\begin{equation}
{L_i} = ||{x_i}|{|^{ - {\alpha _i}}},
\label{eq2}
\tag{2}
\end{equation}
where $\left\| {{x_i}} \right\|$ denotes the distance between $SBS_i$ and UE$_0$, ${\alpha _i}$ is the path loss exponent of the link between $SBS_i$ and UE$_0$. The path loss in real-world environments is affected by electromagnetic radiation, atmospheric environments, weather conditions, obstacle distribution, and diffraction and scattering effects. Considering the fractal characteristics of cellular coverage\cite{X15}, the path loss exponents are usually assumed to be different in different links of 5G small-cell networks. In this paper, the path loss exponents ${\alpha _i}$ ${(i=1,2,3,...)}$ of the links between UE$_0$ and $SBS_i$ ${(i=1,2,3,...)}$ are assumed to be independent identically distributed ${(i.i.d.)}$ random variables. As is well known that the outdoor path loss exponent is larger than its indoor counterpart. The minimum indoor path loss exponent is measured to be 1.9, while its outdoor counterpart  is measured to be 3.1$\sim$4.7 at 900 MHz\cite{A9}. In this case, the path loss exponent of the link between UE$_0$ and $SBS_i$ is assumed to follow a Gamma distribution, i.e., ${\alpha _i}$$\sim$$Gamma(9,0.5)$. The value of the path loss exponent ${\alpha _i}$ is in the interval [2,5.5] with a probability of 0.75. The probability density function (PDF) of the path loss exponent ${\alpha _i}$ is expressed as
\begin{equation}
{f}\left( {\alpha ,m,n} \right) = \frac{{{m^n}}}{{\Gamma (n)}}{\alpha ^{n - 1}}{e^{ - m\alpha }},\alpha  > 0,
\label{eq3}
\tag{3}
\end{equation}
with $m$=0.5 and $n$=9.
\section{Average Achievable Rates with SBS Cooperation Strategies}
\label{sec3}
In order to increase the data rates of UEs at the coverage edge of a small-cell, SBS cooperation strategies are  resorted to. In this section, the general results of the average achievable rate at UE$_0$ with two common SBS cooperation strategies, namely the strategy with a distance constraint and the strategy with a received signal power constraint, are derived in the fractal small-cell networks. The average achievable rate denotes the achievable maximum data rate of the network, which is expressed as
\begin{equation}
\tau  = W{\mathbb{E}}\left[ {\ln \left( {1 + SINR} \right)} \right],
\label{eq4}
\tag{4}
\end{equation}
where $W$ is the bandwidth assigned to UE$_0$, and $SINR$ is the signal-to-interference-plus-noise ratio (SINR) at UE$_0$ adopting SBS cooperation strategies. ${\mathbb{E}}\left[  \cdot  \right]$ is an expectation operation.
\subsection{Average achievable rate adopting the SBS cooperation strategy with distance constraint}
\label{sec3-1}
In the traditional isotropic path loss model, the path loss decreases with the decrease of the distance between a UE and a SBS. Thus, SBSs closer to the UE offer a better channel fading. In the SBS cooperation strategy with a distance constraint, selecting cooperative SBSs is based on the distances between the UE$_0$ and SBSs. The SBS cooperation strategy with the distance constraint is configured that $K$ SBSs closest to UE$_0$ cooperatively transmit the same data to UE$_0$, which is expressed as
\begin{equation}
{\Theta _D} = \left\{ {{x_k} \in {\Phi _B}|\left\| {{{{x_k}}}} \right\| \le \left\| {{{x_K}}} \right\|} \right\},
\label{eq5}
\tag{5}
\end{equation}
where ${\Theta _D}$ denotes the set of cooperative SBSs, ${\left\| {{{x_i}}} \right\|}$ is the distance between $SBS_i$ and UE$_0$, satisfying $\left\| {{{x_1}}} \right\| < \left\| {{x_2}} \right\| < ... < \left\| {{x_k}} \right\| < ...$.
The desired received signal power at UE$_0$ from the cooperative SBSs is given by
\begin{equation}
{P_D} = \sum\limits_{i = 1}^K {{P_s}{h_i}} {\left\| {{x_i}} \right\|^{ - {\alpha _i}}}.
\label{eq6}
\tag{6}
\end{equation}
The interference power aggregated at UE$_0$ is expressed as
\begin{equation}
{I_D} = \sum\limits_{j = K + 1}^\infty  {{P_s}{h_j}} {\left\| {{x_j}} \right\|^{{\rm{ - }}{\alpha _j}}}.
\label{eq7}
\tag{7}
\end{equation}
Furthermore, the average achievable rate with the anisotropic path loss model adopting the distance constraint is derived by
\[\begin{array}{l}
{\tau _D} = W \cdot {{\mathbb{E}}_{{P_D},{I_D}}}\left[ {\ln \left( {1 + \frac{{{P_D}}}{{{I_D} + {\sigma ^2}}}} \right)} \right]\\
\mathop  = \limits^{(a)} W \cdot \int_0^\infty  {\frac{{{e^{^{ - {\sigma ^2}s}}}}}{s}} {\mathcal{L}_{{I_D}}}(s)\left[ {1 - {\mathcal{L}_{{P_D}}}(s)} \right]ds,
\end{array}\,\tag{8}\]
where step (a) utilizes the transfer formula of $\ln \left( {1 + x} \right) = \int_0^\infty  {\frac{1}{z}} \left( {1 - {e^{ - xz}}} \right){e^{ - z}}dz,x > 0$\cite{K16}, ${L_{{P_D}}}(s)$  and ${L_{{I_D}}}(s)$ are the Laplace transforms of the desired received signal power ${P_D}$ and the interference power ${I_D}$, respectively. The Laplace transform of the desired received signal power ${P_D}$ is calculated by
\[\begin{array}{l}
{\mathcal{L}_{{P_D}}}(s) = {{\mathbb{E}}_{{P_D}}}\left[ {\exp \left( { - s{P_D}} \right)} \right]\\
 = {{\mathbb{E}}_{{\Theta_D},\left\{{h_i}\right\},\left\{{\alpha _i}\right\}}}\left[ {\prod\limits_{i = 1}^K {\exp \left( { - s{P_s}{h_i}{{\left\| {{x_i}} \right\|}^{ - {\alpha _i}}}} \right)} } \right]\\
 = {{\mathbb{E}}_{\Theta_D}}\left[ {\prod\limits_{i = 1}^K {{{\mathbb{E}}_{{h},{\alpha}}}\left[ {\exp \left( { - s{P_s}{h}{{\left\| {{x}} \right\|}^{ - {\alpha}}}} \right)} \right]} } \right]\\
 = {{\mathbb{E}}_{\Theta_D}}\left[ {\prod\limits_{i = 1}^K {{{\mathbb{E}}_{{\alpha }}}\left( {{\mathcal{L}_{{h}}}\left( {s{P_s}{{\left\| {{x_i}} \right\|}^{ - {\alpha }}}} \right)} \right)} } \right]\\
\mathop  = \limits^{(a)} {{\mathbb{E}}_{\Theta_D}}\left[ {\prod\limits_{i = 1}^K {{{\mathbb{E}}_{{\alpha }}}\left[ {\frac{1}{{s{P_s}{{\left\| {{x_i}} \right\|}^{ - {\alpha}}} + 1}}} \right]} } \right],
\end{array}\,\tag{9}\]
where step (a) utilizes the Laplace transform of the exponent function $\mathcal{L}(\beta )=\int_0^\infty{{e^{ - t}}{e^{ - \beta t}}}dt=\frac{1}{{\beta  + 1}}$, and submit $\beta=s{P_s}{\left\| {{{\rm{x}}_i}} \right\|^{ - {\alpha}}}$ into the equation. Based on the probability generating functional (PGFL) of Poisson point processes\cite{S11} ${{\mathbb{E}}_\Phi }\left[ {\prod\limits_{x \in \Phi } {f(x)} } \right] = \exp \left( { - \lambda \int\limits_{{R^2}} {\left( {1 - f(x)} \right)dx} } \right)$, the Laplace transform of the desired received signal power ${P_D}$ is further derived by
\[\begin{array}{l}
{\mathcal{L}_{{P_D}}}(s)\\
 = \exp \left( { - 2\pi {\lambda _B}\int_0^{\left\| {{x_K}} \right\|} {\left( {1 - {{\mathbb{E}}_\alpha }\left[ {\frac{1}{{s{P_s}{{\left\| {x} \right\|}^{ - \alpha }} + 1}}} \right]} \right)\left\| x \right\|d\left\| x \right\|} } \right).
\end{array}\,\tag{10}\]
The Laplace transform of the interference power ${I_D}$ is calculated in the same manner as
\[\begin{array}{l}
{\mathcal{L}_{{I_D}}}(s)\\
 = \exp \left( { - 2\pi {\lambda _B}\int_{\left\| {{x_{K + 1}}} \right\|}^\infty  {\left( {1 - {{\mathbb{E}}_\alpha }\left[ {\frac{1}{{s{P_s}{{\left\| {x} \right\|}^{ - \alpha }} + 1}}} \right]} \right)\left\| x \right\|d\left\| x \right\|} } \right).
\end{array}\,\tag{11}\]

Submitting (10) and (11) into (8), the average achievable rate with the anisotropic path loss model adopting the distance constraint is expressed as
\[\begin{array}{l}
{\tau _D}\\
{\rm{ = }}W \cdot \int_0^\infty  {\int_0^\infty  {\int_0^\infty  {\frac{{{e^{^{ - {\sigma ^2}s}}}}}{s}} \left\{ {\exp \left( { - 2\pi {\lambda _B}\int_0^\infty  {\int_{\left\| {{x_{K + 1}}} \right\|}^\infty  {} } } \right.} \right.} } \\
\left( {1 - \left( {\frac{1}{{s{P_s}{{\left\| x \right\|}^{ - \alpha }} + 1}}} \right)} \right)\left. {{f_{{\alpha _i}}}(\alpha )d\left\| x \right\|d{\alpha _{\rm{i}}}} \right)\\
 \times \left[ {1 - \exp \left( { - 2\pi {\lambda _B}\int_0^\infty  {\int_0^{\left\| {{x_K}} \right\|} {\left( {1 - \left( {\frac{1}{{s{P_s}{{\left\| x \right\|}^{ - \alpha }} + 1}}} \right)} \right)} } } \right.} \right.\\
\left. {\left. {\left. {{f_{{\alpha _i}}}(\alpha )d\left\| x \right\|d{\alpha _{\rm{i}}}} \right)} \right]} \right\}dsf\left( {\left\| {{x_K}} \right\|,\left\| {{x_{K + 1}}} \right\|} \right)d\left\| {{x_K}} \right\|d\left\| {{x_{K + 1}}} \right\|,
\end{array}\,\tag{12}\]
where $f\left( {\left\| {{x_K}} \right\|,\left\| {{x_{K + 1}}} \right\|} \right)$ is the joint PDF of $\left\| {{x_K}} \right\|$ and $\left\| {{x_{K + 1}}} \right\|$, which is expressed as \cite{D10}
\[\begin{array}{l}
f(\left\| {{x_K}} \right\|,\left\| {{x_{K + 1}}} \right\|) = \\
\frac{{{{\left( {\pi {\lambda _B}} \right)}^{K + 1}}}}{{\left( {K - 1} \right)!}}{e^{ - \pi {\lambda _B}{{\left\| {{x_{K + 1}}} \right\|}^2}}}{\left\| {{x_K}} \right\|^{2K - 1}}\left\| {{x_{K + 1}}} \right\|.
\end{array}\,\tag{13}\]

\subsection{Average achievable rate adopting the SBS cooperation strategy with received signal power constraint}
\label{sec3-2}
The SBS cooperation strategy with the received signal power constraint is that a subset of the total ensemble of SBSs cooperate by jointly transmitting the same data to UE$_0$. The $SBS_i$ located at $x_i$ belongs to the cooperative set ${\Theta _P}$ of UE$_0$ only if ${{P_s}{h_i}||{x_i}|{|^{ - {\alpha _i}}} \ge T}$, where $T$ is the received signal power threshold at UE$_0$. Thus, the set of the cooperative SBSs with the received signal power constraint for UE$_0$ is
\begin{equation}
{\Theta _P} = \left\{ {{x_i} \in {\Phi _B}|{P_s}{h_i}||{x_i}|{|^{ - {\alpha _i}}} \ge T} \right\}.
\label{eq14}
\tag{14}
\end{equation}
Furthermore, the desired received signal power with the received signal power constraint at UE$_0$ from the cooperative SBSs is given by
\begin{equation}
{P_P} = \sum\limits_{{x_i} \in {\Phi _B}} {{P_s}{h_i}{{\left\| {{x_i}} \right\|}^{ - {\alpha _i}}} \cdot 1\left( {{P_s}{h_i}{{\left\| {{x_i}} \right\|}^{ - {\alpha _i}}} \ge T} \right)},
\label{eq15}
\tag{15}
\end{equation}
where $1\left(  \cdot  \right)$ is an indicator function. The interference power with the received signal power constraint aggregated at UE$_0$ is expressed as
\begin{equation}
{I_P} = \sum\limits_{{x_j} \in {\Phi _B}} {{P_s}{h_j}{{\left\| {{x_j}} \right\|}^{ - {\alpha _j}}} \cdot 1\left( {{P_s}{h_j}{{\left\| {{x_j}} \right\|}^{ - {\alpha _j}}} < T} \right)}.
\label{eq16}
\tag{16}
\end{equation}
The Laplace transforms of the desired received signal power ${P_P}$ and the interference power ${I_P}$ with the received signal power constraint are calculated by
\[\begin{array}{l}
{\mathcal{L}_{{P_P}}}(s) = {{\mathbb{E}}_{{P_P}}}\left[ {\exp \left( { - s{P_P}} \right)} \right]\\
{\rm{ = }}{{\mathbb{E}}_{\left\{ {{\Phi _B}} \right\},\left\{ {{h_i}} \right\},\left\{ {{\alpha _i}} \right\}}}\left[ {\exp \left( { - s\sum\limits_{{x_i} \in {\Phi _B}} {{P_s}{h_i}{{\left\| {{x_i}} \right\|}^{ - {\alpha _i}}}} } \right.} \right.\\
\left. {\left. { \times 1\left( {{P_s}{h_i}{{\left\| {{x_i}} \right\|}^{ - {\alpha _i}}} \ge T} \right)} \right)} \right]\\
 = {{\mathbb{E}}_{\left\{ {{\Phi _B}} \right\},\left\{ {{h_i}} \right\},\left\{ {{\alpha _i}} \right\}}}\left[ {\prod\limits_{{x_i} \in {\Phi _B}} {\exp \left( { - s\left( {{P_s}{h_i}{{\left\| {{x_i}} \right\|}^{ - {\alpha _i}}}} \right)} \right.} } \right.\\
\left. {\left. {\left. { \times 1\left( {{P_s}{h_i}{{\left\| {{x_i}} \right\|}^{ - {\alpha _i}}} \ge T} \right)} \right)} \right)} \right]\\
 = \exp \left( { - 2\pi {\lambda _B}{{\mathbb{E}}_{h,\alpha }}\left[ {{\kappa _1}(r,h,\alpha )} \right]} \right),
\end{array}\,\tag{17}\]
and
\[\begin{array}{l}
{\mathcal{L}_{{I_P}}}(s) = {{\mathbb{E}}_{{I_P}}}\left[ {\exp \left( { - s{I_P}} \right)} \right]\\
{\rm{ = }}{{\mathbb{E}}_{\left\{ {{\Phi _B}} \right\},\left\{ {{h_j}} \right\},\left\{ {{\alpha _j}} \right\}}}\left[ {\exp \left( { - s\sum\limits_{{x_j} \in {\Phi _B}} {{P_s}{h_j}{{\left\| {{x_j}} \right\|}^{ - {\alpha _j}}}} } \right.} \right.\\
\left. {\left. { \times 1\left( {{P_s}{h_j}{{\left\| {{x_j}} \right\|}^{ - {\alpha _j}}} < T} \right)} \right)} \right]\\
 = {{\mathbb{E}}_{\left\{ {{\Phi _B}} \right\},\left\{ {{h_j}} \right\},\left\{ {{\alpha _j}} \right\}}}\left[ {\prod\limits_{{x_j} \in {\Phi _B}} {\exp \left( { - s\left( {{P_s}{h_j}{{\left\| {{x_j}} \right\|}^{ - {\alpha _j}}}} \right)} \right.} } \right.\\
\left. {\left. {\left. { \times 1\left( {{P_s}{h_j}{{\left\| {{x_j}} \right\|}^{ - {\alpha _j}}} < T} \right)} \right)} \right)} \right]\\
 = \exp \left( { - 2\pi {\lambda _B}{{\mathbb{E}}_{h,\alpha }}\left[ {{\kappa _2}(r,h,\alpha )} \right]} \right),
\end{array}\,\tag{18}\]
where${\kappa _1}(r,h,\alpha ) = \int_0^{{{\left( {\frac{{{P_s}h}}{T}} \right)}^{\frac{1}{\alpha }}}} {1 - \exp \left( { - s{P_s}h{r^{ - \alpha }}} \right)rdr}$, and ${\kappa _2}(r,h,\alpha ) = \int_{{{\left( {\frac{{{P_s}h}}{T}} \right)}^{\frac{1}{\alpha }}}}^\infty  {1 - \exp \left( { - s{P_s}h{r^{ - \alpha }}} \right)rdr}$.
Submitting (17)and (18) into (8), the average achievable rate of the SBS cooperation strategy with the received signal power constraint ${\tau _P}$  is expressed as
\[\begin{array}{l}
{\tau _P} = \int_0^\infty  {\left\{ {\left[ {\exp \left( { - 2\pi {\lambda _B}} \right.} \right.} \right.} \int {\int {{e^{{h_j}}}} } {f_{{\alpha _j}}}(\alpha ){\kappa _2}(r,{h_j},{\alpha _j})\\
\left. {\left. {d{\alpha _j}d{h_j}} \right)} \right] \times \left[ {1{\rm{ - }}\exp \left( { - 2\pi {\lambda _B}\int {\int {{{\rm{e}}^{{h_i}}}} } } \right.} \right.{f_{{\alpha _i}}}(\alpha )\\
\left. {\left. {\left. {{\kappa _1}(r,{h_i},{\alpha _i})d{\alpha _i}d{h_i}} \right)} \right]} \right\}\frac{{{e^{^{ - {\sigma ^2}s}}}}}{s}ds \cdot W,
\end{array}\,\tag{19}\]
where ${f_{{\alpha _j}}}(\alpha )$ is the PDF of the path loss exponent ${\alpha _i}$.

What's more, the number of cooperative SBSs of UE$_0$ is calculated by
\begin{equation}
{N_C} = \sum\limits_{{x_i} \in {\Phi _B}} {1\left( {{P_s}{h_i}{{\left\| {{x_i}} \right\|}^{ - {\alpha _i}}} \ge T} \right)}.
\label{eq20}
\tag{20}
\end{equation}
The average number of cooperative SBSs is further calculated by
\[\begin{array}{l}
{\mathbb{E}}\left[ {{N_C}} \right]\\
 = {{\mathbb{E}}_{{\Phi _B},\left\{ {{h_i}} \right\},\left\{ {{\alpha _i}} \right\}}}\left[ {\sum\limits_{{x_i} \in {\Phi _B}} {1\left( {{P_s}{h_i}{{\left\| {{x_i}} \right\|}^{ - {\alpha _i}}} \ge T} \right)} } \right]\\
\mathop  = \limits^{\left( a \right)} {{\mathbb{E}}_{\left\{ {{h_i}} \right\},\left\{ {{\alpha _i}} \right\}}}\left[ {2\pi {\lambda _B}\int_0^\infty  {1\left( {{P_s}{h_i}{{\left\| r \right\|}^{ - {\alpha _i}}} \ge T} \right)} rdr} \right]\\
 = {{\mathbb{E}}_{\left\{ {{h_i}} \right\},\left\{ {{\alpha _i}} \right\}}}\left[ {2\pi {\lambda _B}\int_0^\infty  {1\left( {\left\| r \right\| \le {{\left( {\frac{{{P_s}{h_i}}}{T}} \right)}^{\frac{1}{{{\alpha _i}}}}}} \right)} rdr} \right]\\
 = {{\mathbb{E}}_{\left\{ {{h_i}} \right\},\left\{ {{\alpha _i}} \right\}}}\left[ {\pi {\lambda _B}{{\left( {\frac{{{P_s}{h_i}}}{T}} \right)}^{\frac{2}{{{\alpha _i}}}}}} \right]\\
 = \pi {\lambda _B}\int {{{\left( {\frac{{{P_s}}}{T}} \right)}^{\frac{2}{{{\alpha _i}}}}}} \left( {\int {{e^{{h_i}}}{h_i}^{\frac{2}{{{\alpha _i}}}}d{h_i}} } \right)f({\alpha _i})d{\alpha _i}.
\end{array}\,\tag{21}\]
Letting $K = {\mathbb{E}}\left[ {{N_C}} \right]$, the average achievable rate with distance constraint considering the same number of cooperative SBSs is expressed as
\[\begin{array}{l}
{\tau _D}\left( K \right){\rm{ = }}W \cdot \int_0^\infty  {\int_0^\infty  {\int_0^\infty  {\frac{{{e^{^{ - {\sigma ^2}s}}}}}{s}} \left\{ {\exp \left( { - 2\pi {\lambda _B}\int_0^\infty  {\int_{\left\| {{x_{K + 1}}} \right\|}^\infty  {} } } \right.} \right.} } \\
\left( {1 - \left( {\frac{1}{{s{P_s}{{\left\| x \right\|}^{ - \alpha }} + 1}}} \right)} \right)\left. {{f_{{\alpha _i}}}(\alpha )d\left\| x \right\|d{\alpha _{\rm{i}}}} \right)\\
 \times \left[ {1 - \exp \left( { - 2\pi {\lambda _B}\int_0^\infty  {\int_0^{\left\| {{x_K}} \right\|} {\left( {1 - \left( {\frac{1}{{s{P_s}{{\left\| x \right\|}^{ - \alpha }} + 1}}} \right)} \right)} } } \right.} \right.\\
\left. {\left. {\left. {{f_{{\alpha _i}}}(\alpha )d\left\| x \right\|d{\alpha _{\rm{i}}}} \right)} \right]} \right\}dsf\left( {\left\| {{x_K}} \right\|,\left\| {{x_{K + 1}}} \right\|} \right)d\left\| {{x_K}} \right\|d\left\| {{x_{K + 1}}} \right\|,
\end{array}\,\tag{22a}\]
with
\[\begin{array}{l}
f(\left\| {{x_K}} \right\|,\left\| {{x_{K + 1}}} \right\|) = \\
\frac{{{{\left( {\pi {\lambda _B}} \right)}^{K + 1}}}}{{\left( {K - 1} \right)!}}{e^{ - \pi {\lambda _B}{{\left\| {{x_{K + 1}}} \right\|}^2}}}{\left\| {{x_K}} \right\|^{2K - 1}}\left\| {{x_{K + 1}}} \right\|.
\end{array}\,\tag{22b}\]
The increment in the average achievable rate compared the received signal power constraint with the distance constraint considering same number of cooperative SBSs is given by
\begin{equation}
g = \frac{{{\tau _P}({\mathbb{E}}\left[ {{N_C}} \right]) - {\tau _D}({\mathbb{E}}\left[ {{N_C}} \right])}}{{{\tau _D}({\mathbb{E}}\left[ {{N_C}} \right])}}.
\label{eq23}
\tag{23}
\end{equation}

\section{Network Energy Efficiency}
\label{sec4}

The average achievable rate can be improved by adopting SBS cooperation strategies. However, more SBS resources are required to transmit data in the cooperative small-cell networks. It is important to study the energy efficiency of networks for both economical and environmental considerations\cite{X11}. In this section, the energy efficiency of the SBS cooperation strategy with the received signal power constraint is analyzed. To simplify the calculation, a linear approximation of the SBS power model\cite{G12} is taken into consideration, which is expressed as
\begin{equation}
{P_{SBS}} = {P_0} + {N_{UE}}{P_s}{\Delta _p},
\label{eq24}
\tag{24}
\end{equation}
where ${P_0}$ is the power consumption at the minimum non-zero output power, and ${\Delta _p}$ is the slope of the loaded-dependent power consumption. ${N_{UE}}$ denotes the average number of UEs serviced by a SBS.
The average number of UEs served by a SBS can be calculated, which is expressed as
\[\begin{array}{l}
{\mathbb E}\left[ {{N_{UE}}} \right]= \pi {\lambda _U}\int {{{\left( {\frac{{{P_s}}}{T}} \right)}^{\frac{2}{{{\alpha _i}}}}}} \left( {\int {{e^{{h_i}}}{h_i}^{\frac{2}{{{\alpha _i}}}}d{h_i}} } \right){f_{{\alpha _i}}}(\alpha )d{\alpha _i}.
\end{array}\,\tag{25}\]
The sum rate of all the UEs in the fractal small-cell networks is expressed as
\[\begin{array}{l}
{\tau _{sum}} = \sum\limits_{{y_u} \in {\Phi _U}}^{}{{\mathbb{E}}\left[ {\ln \left( {1 + SIN{R_u}} \right)} \right]} \\
{\rm{ = }}S \cdot {\lambda _U} \cdot {\mathbb{E}}\left[ {\ln \left( {1 + SINR} \right)} \right]\\
{\rm{ = }}W \cdot \int_0^\infty  {\left\{ {\left[ {\exp \left( { - 2\pi {\lambda _B}} \right.} \right.} \right.} \int {\int {{e^{{h_j}}}} } f({\alpha _j})\kappa_2 (r,{h_j},{\alpha _j})\\
\left. {\left. {d{\alpha _j}d{h_j}} \right)} \right] \times \left[ {1{\rm{ - }}\exp \left( { - 2\pi {\lambda _B}\int {\int {{{\rm{e}}^{{h_i}}}} } } \right.} \right.f({\alpha _i})\\
\left. {\left. {\left. {\kappa_1 (r,{h_i},{\alpha _i})d{\alpha _i}d{h_i}} \right)} \right]} \right\}\frac{{{e^{^{ - {\sigma ^2}s}}}}}{s}ds \cdot {\lambda _U} \cdot S,
\end{array}\,\tag{26}\]
where $S$ is the area of interest, ${SINR_u}$ is the SINR of ${UE_u}$ considering the SBS cooperation strategy with the received signal power constraint. Furthermore, the total SBS power consumption of the entire fractal small-cell network is given as
\[\begin{array}{l}
{P_{sum}} = \sum\limits_{{x_i} \in {\Phi _B}} {{P_{SBS}}} = \sum\limits_{{x_i} \in {\Phi _B}} {{P_0} + {N_{UE}}{P_s}{\Delta _p}} )\\
 = S \cdot {\lambda _B} \cdot \left( {P_0} + {N_{UE}}{P_s}{\Delta _p} \right).
\end{array}\,\tag{27}\]
Since the network energy efficiency $\eta$ is defined as a ratio of the average rate to the total SBS power consumption, it can be calculated as
\[\begin{array}{l}
\eta  = \frac{{{\tau _{sum}}}}{{{P_{sum}}}}\\
= \int_0^\infty  {\left\{ {\left[ {\exp \left( { - 2\pi {\lambda _B}} \right.} \right.} \right.} \int {\int {{e^{{h_j}}}} } {f_{{\alpha _j}}}(\alpha ){\kappa _2}(r,{h_j},{\alpha _j}){d{\alpha _j}d{h_j}}\\
 \times \left[ {1{\rm{ - }}\exp \left( { - 2\pi {\lambda _B}\int {\int {{{\rm{e}}^{{h_i}}}} } } \right.} \right.{f_{{\alpha _i}}}(\alpha )\left. {\left. {\left. {{\kappa _1}(r,{h_i},{\alpha _i})d{\alpha _i}d{h_i}} \right)} \right]} \right\}\\\frac{{{e^{^{ - {\sigma ^2}s}}}}}{s}ds \cdot {\lambda _U}\cdot W/{\left({\left( {P_0} + {N_{UE}}{P_s}{\Delta _p} \right) \cdot {\lambda _B}}\right)}.
\end{array}\,\tag{28}\]

\section{Simulation Results and Discussion}
\label{sec5}

In this section, simulation results on the average achievable rate and network energy efficiency of the fractal small-cell networks are analyzed. In the following analysis, some default parameters are configured as: ${\lambda _U} = \frac{1}{{300\pi }}$, $W=1$, ${P_s} = 0.13W$, ${\Delta _p} = 4$, ${P_0} = 2.5W$, and ${\sigma ^2} =  - 95dBm$ \cite{Y13}.

Fig. 2 plots the average number of users per SBS ${N_{UE}}$ and the number of cooperative SBSs ${N_C}$ with respect to received signal power thresholds considering various intensities of SBS cooperation. Solid lines represent ${N_{UE}}$, which refers to the left vertical axis. Dotted lines represent ${N_C}$, which refers to the right vertical axis. When the intensity of the SBSs is fixed, both ${N_{UE}}$ and ${N_C}$ decrease with the increase of the received signal power threshold. When the received signal power threshold is fixed, ${N_C}$ increases with the increase of the intensity of SBSs from $\frac{1}{{{{100}^2}\pi }}$ to $\frac{1}{{{{50}^2}\pi }}$. ${N_{UE}}$ increases with the increase of the intensity of UEs from $\frac{1}{{{500}\pi }}$ to $\frac{1}{{{100}\pi }}$.
\begin{figure}
\setlength{\abovecaptionskip}{0.cm}
\setlength{\belowcaptionskip}{-0.cm}
  \centering
  \includegraphics[width=7cm,draft=false]{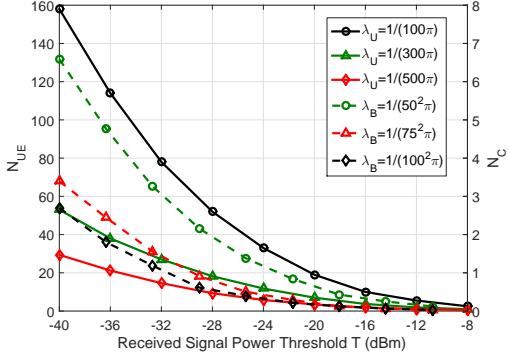}\\
  \caption{\small The average number of users per SBS serves ${N_{UE}}$ and the number of cooperative SBSs ${N_C}$ with respect to different received signal power threshold $T$ adopting SBS cooperation strategy with received signal power constraint.}
  \label{fig2}
\end{figure}

Fig. 3 illustrates the average achievable rate with respect to the number of cooperative SBSs $K$ considering different intensities of SBSs when the SBS cooperation strategy with distance constraint is adopted. When the intensity of SBSs is fixed, the average achievable rate increases with the increase of $K$ from 1 to 6. When the number of cooperative SBSs $K$ is fixed, the average achievable rate decreases with the increase of the intensity of SBSs from $\frac{1}{{{{100}^2}\pi }}$ to $\frac{1}{{{{20}^2}\pi }}$. In the case adopting the distance constraint, the interference increases more than the desired received signal power with the increase of the intensity of SBSs since the interference includes higher received signal powers than the desired received signal power of coordinated SBSs. The SINR is reduced by increasing the intensity, so that the average achievable rate decreases with the increase of the intensity of SBSs.
\begin{figure}
\setlength{\abovecaptionskip}{0.cm}
\setlength{\belowcaptionskip}{-0.cm}
  \centering
 \includegraphics[width=7cm,draft=false]{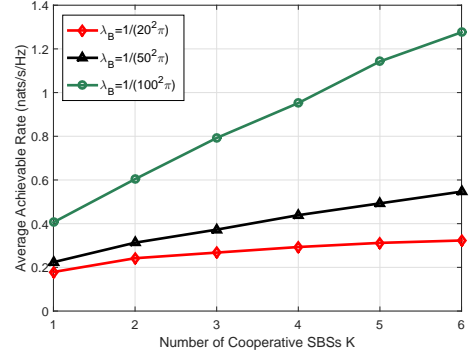}\\
 \caption{\small The average achievable rate with respect to the number of cooperative SBSs $K$ considering different intensities of SBSs when the SBS cooperation strategy with distance constraint is adopted.}
 \label{fig3}
\end{figure}

Fig. 4 shows the average achievable rate with respect to the received signal power threshold considering different intensities of SBSs when the SBS cooperation strategy with the received signal power constraint is adopted. When the intensity of the SBSs is fixed, the average achievable rate decreases with the increase of the received signal power threshold, since the number of cooperative SBSs becomes smaller. When the received signal power threshold is fixed, the average achievable rate increases with the increase of the intensity of the SBSs from $\frac{1}{{{{100}^2}\pi }}$ to $\frac{1}{{{{20}^2}\pi }}$.
\begin{figure}
  \centering
  \includegraphics[width=7cm,draft=false]{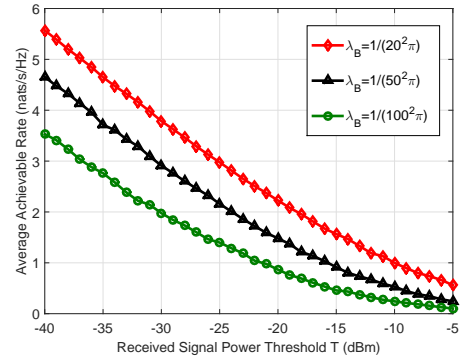}\\
  \caption{\small The average achievable rate with respect to the received signal power threshold $T$ considering different intensities of the SBSs when the SBS cooperation strategy with received signal power constraint is adopted.}
  \label{fig4}
\end{figure}

\begin{table}
\caption{Comparison of average achievable rates between distance constraint and received signal power constraint ($\lambda_B=\frac{1}{{{{50}^2}\pi }}$)}
\begin{center}
\begin{tabu} to 0.4\textwidth{|X[c]|X[3,c]|X[3,c]|X[3,c]|X[3,c]|}
\hline
$K$  &$T$              &Rate A     &Rate B & $g$\\
\hline
1          &-22$dBm$           &0.2237    &1.722   &7.00 \\
\hline
2         &-28$dBm$           &0.3129    &2.606    &7.32\\
\hline
3         &-32$dBm$           &0.3727   &3.291    &7.83\\
\hline
4        &-35$dBm$          &0.4389     &3.713    &7.45\\
\hline
5         &-37$dBm$          &0.4929   &4.138    &7.39\\
\hline
6          &-39$dBm$          &0.5466   &4.478  &7.20\\
\hline
\end{tabu}
\end{center}
\end{table}

Comparing the two cooperation strategies, it is found that the average achievable rate of the cooperation strategy with the received signal power constraint is much more higher than that of the cooperation strategy with a distance constraint when the numbers of cooperative SBSs of two strategies are same. Furthermore, the rate increment comparing two strategies can become larger than seven shown in Table I. Rate A denotes the average achievable rate adopting SBS strategy with a distance constraint, and Rate B denotes the average achievable rate adopting SBS strategy with the received signal power constraint. Therefore, in the practical scenarios, the cooperation strategy with the received signal power constraint can provide better rate performance than the strategy with a distance constraint.

Fig. 5 depicts the network energy efficiency of fractal small-cell networks with respect to the received signal power threshold $T$ considering different intensities of SBSs. When the received signal power threshold is fixed, the network energy efficiency decreases with the increase of the intensity of SBSs from $\frac{1}{{{{100}^2}\pi }}$ to $\frac{1}{{{{50}^2}\pi }}$. When the intensity of SBSs is fixed, it can be found that the network energy efficiency increases first and then decreases with the increase of the received signal power threshold. The maximum network energy efficiency can be achieved by adjusting the threshold.
\begin{figure}
\setlength{\abovecaptionskip}{0.cm}
\setlength{\belowcaptionskip}{-0.cm}
  \centering
  \includegraphics[width=7cm,draft=false]{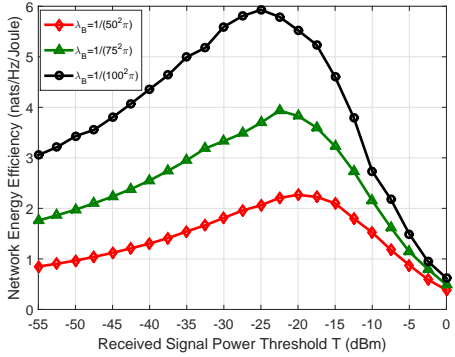}\\
  \caption{\small The network energy efficiency of fractal small-cell networks with respect to the received signal power threshold $T$ considering different intensities of SBSs.}
  \label{fig5}
\end{figure}

\section{Conclusion}
\label{sec6}
In this paper, downlink SBS cooperation strategies with the distance and the received signal power constraints were analyzed based on the anisotropic path loss model. The average achievable rate and the network energy efficiency were derived for fractal small-cell networks. Simulation results show that the average achievable rate with the received signal power constraint is larger than the rate with the distance constraint in consideration of the same number of cooperative SBSs. What's more, the network energy efficiency of fractal small-cell networks adopting the SBS cooperation strategy with the received signal power constraint was analyzed. The network energy efficiency decreases with the increase of the intensity of SBSs and can achieve a maximum value by adjusting the received signal power threshold. The advantage of increasing the intensity of SBSs is weakened by the SBS cooperation with the received signal power constraint, which indicates a challenge on the deployment and SBS cooperation design for fractal small-cell networks.

\section*{Acknowledgment}
The authors would like to acknowledge the support from the National Natural Science Foundation of China (NSFC) under Grants 61210002, the Fundamental Research Funds for the Central Universities under the Grant 2015XJGH011. This research is partially supported by the EU FP7-PEOPLE-IRSES, project acronym CROWN (Grant no. 610524). This research is supported by the National international Scientific and Technological Cooperation Base of Green Communications and Networks (No. 2015B01008)






\begin{thebibliography}{99}

\bibitem{V1}
G. Mao and B. D. O. Anderson, ``Connectivity of Large Wireless Networks Under A General Connection Model," {\em IEEE Trans. Inf. Theory.}, vol. 59, no. 3, pp. 1761--1772, Mar. 2013.

\bibitem{T2}
X. Ge, S. Tu, G. Mao, et al., ``5G Ultra-Dense Cellular Networks," {\em IEEE Wireless Commun.}, vol. 23, no. 1, pp. 72--79, Feb. 2016.


\bibitem{S2}
S. Akoum and R. Heath, ``Multi-cell coordination: A stochastic geometry approach," {\em Proc. IEEE 13th Int. Workshop Signal Process. Adv. Wireless Commun.}, Jun. 2012, pp. 16--20.


\bibitem{X4}
X. Ge, H. Cheng, G. Mao, et al., ``Vehicular Communications for 5G Cooperative Small-Cell Networks," {\em IEEE Trans. Veh. Technol.}, vol. 65, no. 10, pp. 7882--7894, Oct. 2016.

\bibitem{G5}
G. Nigam, P. Minero and M. Haenggi, ``Coordinated Multipoint Joint Transmission in Heterogeneous Networks," {\em IEEE Trans. Commun.}, vol. 62, no. 11, pp. 4134--4146, Nov. 2014.

\bibitem{W6}
W. Nie, F. C. Zheng, X. Wang, et al., ``User-Centric Cross-Tier Base Station Clustering and Cooperation in Heterogeneous Networks: Rate Improvement and Energy Saving," {\em IEEE J. Sel. Areas Commun.}, vol. 34, no. 5, pp. 1192--1206, May 2016.

\bibitem{G8}
X. Ge, Y. Qiu, J. Chen, et al., ``Wireless fractal cellular networks," {\em IEEE Wireless Commun.}, vol. 23, no. 5, Oct. 2016.

\bibitem{A13}
A. Guo and M. Haenggi, ``Spatial Stochastic Models and Metrics for the Structure of Base Stations in Cellular Networks," {\em IEEE Trans. Wireless Commun.}, vol. 12, no. 11, pp. 5800--5812, Nov. 2013.

\bibitem{G1}
G. Mao and B. D. Anderson, ``Capacity of Large Wireless Networks with Generally Distributed Nodes," {\em IEEE Trans. Wireless Commun.}, vol. 13, no. 3, pp. 1678--1691, Mar. 2014.

\bibitem{A14}
A. Goldsmith, {\em Wireless communications}, Cambridge university press, 2005.

\bibitem{X15}
X. Ge et al., ``Wireless Single Cellular Coverage Boundary Models," {\em IEEE Access}, vol. 4, no. , pp. 3569--3577, 2016.

\bibitem{A9}
A. Bhuvaneshwari, R. Hemalatha and T. Satyasavithri, ``Development of an empirical power model and path loss investigations for dense urban region in Southern India," {\em 2013 IEEE 11th Malaysia International Conference on Communications (MICC)}, Kuala Lumpur, 2013, pp. 500--505.

\bibitem{K16}
K. A. Hamdi, ``Capacity of MRC on Correlated Rician Fading Channels," {\em IEEE Trans. Commun.}, vol. 56, no. 5, pp. 708--711, May 2008.

\bibitem{S11}
S. N. Chiu, D. Stoyan, W. S. Kendall, et al., {\em Stochastic Geometry and Its Applications.} 3rd ed. Hoboken, NJ, USA: Wiley, 2013.

\bibitem{D10}
D. Moltchanov, ``Survey paper: Distance distributions in random networks," {\em Ad Hoc Netw.}, vol. 10, no. 6, pp. 1146--1166, Aug. 2012.

\bibitem{X11}
X. Ge et al., ``Spectrum and Energy Efficiency Evaluation of Two-Tier Femtocell Networks With Partially Open Channels," {\em IEEE Trans. Veh. Technol.}, vol. 63, no. 3, pp. 1306--1319, Mar. 2014.

\bibitem{G12}
G. Auer et al., ``How much energy is needed to run a wireless network?," {\em IEEE Wireless Commun.}, vol. 61, no. 6, pp. 2350--2361, Jun. 2013.

\bibitem{Y13}
X. Ge, J. Ye, Y. Yang, et al., ``User Mobility Evaluation for 5G Small Cell Networks Based on Individual Mobility Model," {\em IEEE J. Sel. Areas Commun.}, vol. 34, no. 3, pp. 528--541, Mar. 2016.


\end{thebibliography}
%

\end{document}